\documentclass{article}
\usepackage{spconf,amsmath,graphicx,hyperref}
\usepackage{booktabs} 
\usepackage{caption}
\usepackage{subcaption}
\usepackage{adjustbox}

\title{Enabling Multi-Species Bird Classification on Low-Power Bioacoustic Loggers}
\name{Stefano Ciapponi$^{1,2}$, Leonardo Mannini$^1$, Jarek Scanferla$^3$, Matteo Anderle$^3$, Elisabetta Farella$^1$ }
\address{Fondazione Bruno Kessler$^1$, University of Trento$^2$, Eurac Research$^3$}

\begin{document}
\maketitle

\begin{abstract}
This paper introduces WrenNet, an efficient neural network enabling real-time multi-species bird audio classification on low-power microcontrollers for scalable biodiversity monitoring. We propose a semi-learnable spectral feature extractor that adapts to avian vocalizations, outperforming standard mel-scale and fully-learnable alternatives. On an expert-curated 70-species dataset, WrenNet achieves up to 90.8\% accuracy on acoustically distinctive species and 70.1\% on the full task. When deployed on an AudioMoth device ($\leq$1MB RAM), it consumes only 77mJ per inference. Moreover, the proposed model is over 16x more energy-efficient compared to Birdnet when running on a Raspberry Pi 3B+. This work demonstrates the first practical framework for continuous, multi-species acoustic monitoring on low-power edge devices.
\end{abstract}


\section{Introduction}

The urgent need for scalable environmental monitoring to address declining biodiversity has made automated acoustic classification a critical alternative to traditional, labor-intensive methods \cite{manzano-rubio_low-cost_2022}. Avian species, as sensitive bioindicators of ecosystem health \cite{bird-indicators}, are a primary focus. While deep learning systems like BirdNET \cite{kahl_birdnet_2021} achieve high accuracy in the lab, their computational demands render them impractical for deployment on severely resource-constrained edge devices (${\leq}1$MB RAM, ${\leq}100$MHz processors).

\begin{figure}[!ht]
    \centering
    \includegraphics[width=1\linewidth]{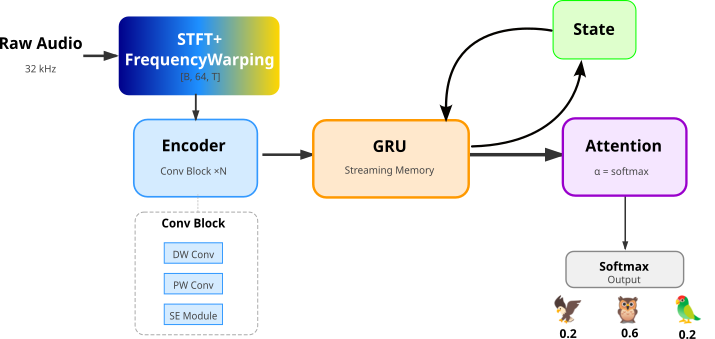}
    \vspace{-6mm}
    \caption{WrenNet processes raw audio through STFT and frequency warping, followed by causal CNN feature extraction, GRU temporal modeling with cross-frame state retention, and attention-based classification.}
    \label{fig:architecture}
\end{figure}

Current hardware solutions range from data-loggers like the AudioMoth \cite{hill_audiomoth_2019} to more capable single-board computers like the Raspberry Pi \cite{delgado-rajo_flexible_2025,treskova_prototyping_2025}. The former lacks on-device processing, while the latter is often too energy-intensive for long-term remote use. While some microcontroller-class solutions exist, they are limited to single-species or binary classification \cite{huang_tinychirp_2024, solomes_efficient_2020}. To our knowledge, no prior work has demonstrated multi-species bird classification on microcontroller hardware.

This gap is critical, as remote deployments require ultra-efficient, battery-powered systems that can operate for extended periods. Continuous on-device classification is highly desirable, even at a slight cost to accuracy, to avoid multi-day retrieval expeditions. This imposes three core signal processing challenges: (1) designing neural architectures that balance accuracy with a tight compute/memory budget; (2) enabling causal processing of long audio with a fixed memory footprint; and (3) developing adaptive feature extraction to capture the wide frequency ranges of avian vocalizations, which fixed mappings like mel-scale filterbanks fail to model effectively.

Our main contributions address these challenges:

\textit{(1) WrenNet}, a streaming-compatible neural architecture using causal convolutions and a unidirectional GRU \cite{gru} for efficient multi-species classification.

\textit{(2) Semi-learnable spectral features} with a parametrized sigmoid-weighted frequency mapping \cite{cai_development_2023, ravanelli_speaker_2018} to adapt resolution to bird vocalizations.

\textit{(3) Comprehensive edge deployment benchmarks} on AudioMoth and Raspberry Pi 3B+, demonstrating practical feasibility.

\textit{(4) An expert-informed alpine bird benchmark} of 70 species, curated with ornithologists \cite{Scanferla2025-us} and annotated with species classification difficulty.
\section{Neural Network Architecture}

The proposed WrenNet architecture addresses computational and memory constraints of avian classification through a hybrid design combining spectral feature extraction with efficient temporal modeling. Unlike conventional approaches processing entire spectrograms through bidirectional networks—requiring substantial memory and precluding streaming applications—our architecture employs causal 1D convolutions with dilated kernels\cite{van2016wavenet} for expanded temporal receptive fields. All convolutions use stride-1 to preserve temporal information density. The convolutional frontend utilizes PhiNet-inspired\cite{paissan_phinets_2022} depthwise separable convolution blocks \cite{howard2017mobilenetsefficientconvolutionalneural}, decomposing standard convolutions into depthwise temporal filtering followed by pointwise channel mixing, with integrated squeeze-and-excitation\cite{squeeze} attention for adaptive channel recalibration.

The encoder employs a modified MatchboxNet\cite{majumdar_matchboxnet_2020} structure with hierarchical skip connections preserving both local temporal details and long-range acoustic structure. The temporal processing stage uses a unidirectional GRU that processes short audio segments sequentially rather than entire recordings, maintaining temporal context between segments through its hidden state. This approach dramatically reduces memory requirements by replacing large spectrogram buffers with compact hidden state representations, enabling streaming processing where arbitrarily long recordings can be processed with fixed, minimal memory overhead. The GRU output undergoes temporal attention-based aggregation to identify discriminative time frames for species classification. The architecture maintains strict causality throughout, ensuring compatibility with streaming deployment scenarios.

A diagram of the architecture is available in Figure \ref{fig:architecture}. The WrenNet configuration employs sigmoid-weighted log-linear frequency mapping with 64 adaptive filter bins, using 512-point FFT with 320-sample hop length and learnable parameters including an 8kHz breakpoint and transition width. The convolutional frontend utilizes PhiNet depthwise separable architecture structured as a 3-layer MatchboxNet with 32 base filters and squeeze-and-excitation attention. For temporal processing, the network employs a unidirectional GRU with 64-dimensional hidden state and temporal attention-based aggregation.
\vspace{-4mm}
\section{Feature Extraction}
\vspace{-2mm}
Mel-scale filterbanks, commonly used in neural networks for audio classification, apply a fixed frequency warping that emphasizes low-frequency resolution (below 1 kHz) while compressing higher frequencies. This design reflects human auditory perception but is suboptimal for avian bioacoustics, where many species exhibit critical spectral signatures at higher bands. Fixed warping functions cannot adapt to diverse spectral distributions across species, leaving neural networks to compensate through generalization rather than receiving appropriately structured inputs. To address this, we propose a differentiable frequency warping function that transitions smoothly from logarithmic to linear scaling via a parametrized sigmoid, enabling gradient-based optimization of spectral resolution as part of an end-to-end differentiable signal processing front-end.

\subsection{Design}

Our method is characterized by two learnable parameters: the \textit{breakpoint} $b$ (transition frequency) and \textit{transition width} $w$ (sharpness). Given normalized frequency coordinates $x_i = \frac{i}{n-1}$ for $i \in \{0,1,\ldots,n-1\}$ and $f_{min}$ and $f_{max}$ the minimum and maximum frequency respectively , we define logarithmic and linear frequency mappings as:

\begin{equation}
    f_{\log}(x) = f_{\min} \left(\frac{f_{\max}}{f_{\min}}\right)^x, \quad f_{\text{linear}}(x) = f_{\min} + x(f_{\max} - f_{\min})
\end{equation}

The parametrized sigmoid weighting function is given by:

\begin{equation}
    S(x; b, w) = \sigma\left(\left(x - \frac{b - f_{\min}}{f_{\max} - f_{\min}}\right) \cdot w\right)    
\end{equation}

where $\sigma(z) = \frac{1}{1 + e^{-z}}$ is the standard sigmoid function.

Our adaptive frequency mapping combines both mappings through a Sigmoid-weighted convex combination:

\begin{equation}
    f(x; b, w) = (1 - S(x; b, w)) f_{\log}(x) + S(x; b, w) f_{\text{linear}}(x)
\end{equation}

where $S(x; b, w) \approx 0$ yields logarithmic mapping and $S(x; b, w) \approx 1$ yields linear mapping.

The discrete FFT bin indices are computed as:

\begin{equation}
    \text{bin}(x) = \text{clip}\left(\left\lfloor\frac{f(x; b, w)}{f_{\text{Nyquist}}} \cdot \frac{n_{\text{fft}}}{2}\right\rfloor, 1, \frac{n_{\text{fft}}}{2} - 2\right)
\end{equation}

where $f_{\text{Nyquist}} = \frac{\text{sr}}{2}$. The limiting cases are: $w \to \infty$ produces a hard switch at frequency $b$, while $w \to 0$ yields uniform blending across all frequencies. 

\begin{figure}[!htbp]
    \centering
    \includegraphics[width=1\linewidth]{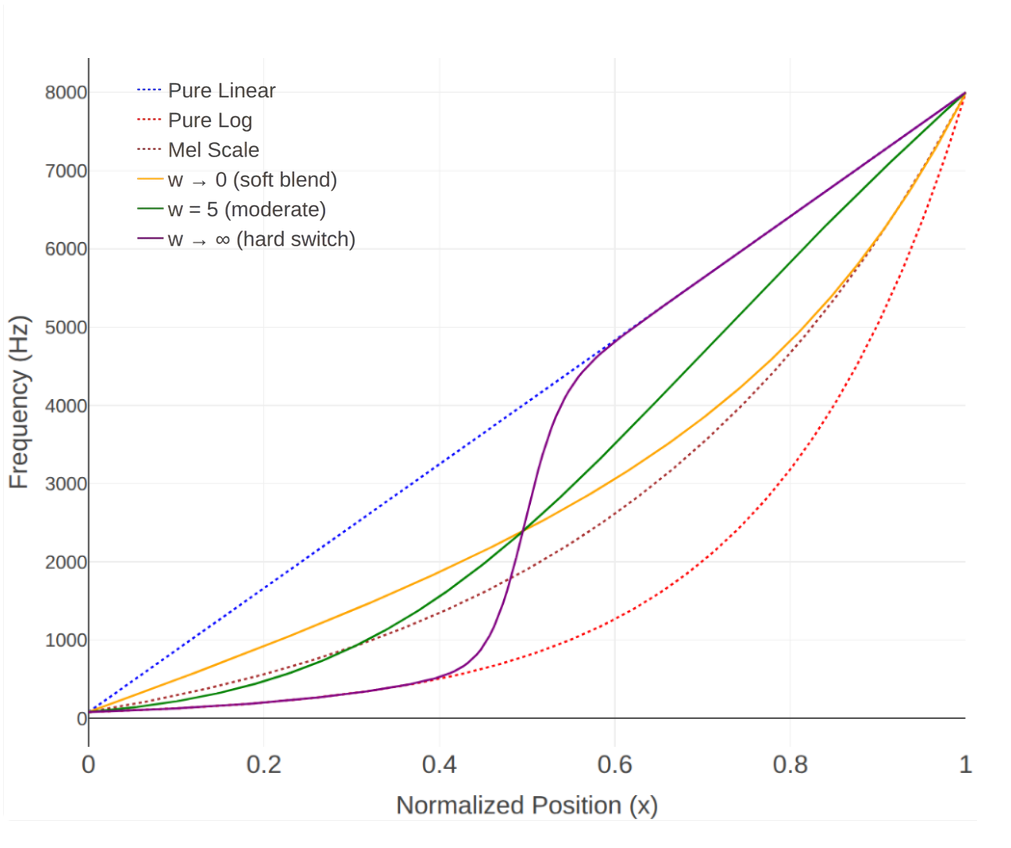}
    \caption{Extremes of the Semi-Learnable filterbanks Frequency mappings.}
    \label{fig:extremes}
\end{figure}

Figure \ref{fig:extremes} showcases differences between different configurations of $w$ with fixed $b=4000$ and how it compares with mel, linear and log mappings. These bin indices define the center frequencies for constructing triangular filterbank filters, following standard mel-filterbank design principles but with our adaptive frequency spacing.

\section{Experimental Setup}


\begin{table*}[!ht]
    \centering
    \small
    \renewcommand{\arraystretch}{1}
    \begin{adjustbox}{width=\linewidth,center}
    \begin{tabular}{lcccccccc}
        \textbf{Configuration} & \textbf{\#Species} & \textbf{Epochs} & \textbf{Test Acc (\%)} & \textbf{F1 (\%)} & \textbf{Student Acc (\%)} & \textbf{BirdNET Acc (\%)} & \textbf{Breakpoint (Hz)} & \textbf{Trans. Width} \\ \hline
        Monoclass (Corvus Corax)     & 1  & 94  & 92.37 & 92.62 & 90.15 & 91.16 & 1955 & 51.79 \\
        Monoclass (Upupa Epops)      & 1  & 45  & 94.71 & 94.67 & 86.35 & 89.45 & 5269 & 132 \\ \hline
        Easy species (semi-learnable)     & 8  & 116 & 90.76 & 90.90 & 89.85 & 91.04 & 1.5 & 6 \\
        \hline
        Semi-learnable (subset) & 8  & 75  & 87.22 & 87.25 & 86.33 & 88.19 & 2741 & 67.79 \\
        Fully learnable (subset baseline)         & 8  & 75  & 83.82 & 83.84 &   -   &   -   & - & - \\ \hline
        Hard species                & 13 & 134 & 77.47 & 77.97 & 73.90 & 82.86 & 1224 & 27.42 \\
        Regulus pair                & 2  & 85  & 85.64 & 85.60 & 83.60 & 87.87 & 1221 & 28.34 \\ \hline
        High-frequency subset       & 5  & 93  & 91.49 & 91.55 & 90.23 & 92.74 & 164 & 8.55 \\
        Low-frequency subset        & 4  & 126 & 91.63 & 91.67 & 85.15 & 91.01 & 237 & 9 \\ \hline
        Full dataset (57k params)   & 70 & 150 & 66.51 & 67.49 &   -   &   -   & 851 & 19 \\
        Full dataset (136k params)  & 70 & 75  & 70.14 & 70.81 &   -   &   -   & 1390 & 33.86 \\ \hline
    \end{tabular}
    \end{adjustbox}
    \caption{Classification performance across different species subsets with learned filter parameters.  The "no\_bird" class was excluded from the Student/Birdnet test set for a fair comparison, as Birdnet does not explicitly model it.}
    \label{tab:results}
\end{table*}

\subsection{Dataset}
Our dataset consists of 70 alpine bird species \cite{Scanferla2025-us} representative of Central European mountain ecosystems. Audio recordings were sourced from Xeno-Canto\footnote{https://xeno-canto.org/} using an automated querying pipeline, resulting in 150,645 downloaded audio files. Raw audio was resampled to 32~kHz (mono) and bandpass filtered between 150~Hz and 16~kHz to remove noise while preserving the spectral range relevant to bird vocalizations.

To extract vocalization segments, we applied adaptive peak detection on amplitude envelopes computed using 50~ms sliding windows with 10~ms hop size. Prominence-based thresholding was used with a minimum inter-peak distance of 1~second to avoid duplicates. Peaks were ranked by prominence, and the top segments were selected using the 75th percentile threshold. If multiple calls were detected, the first was selected; if none were found, a fallback segment was extracted from the initial 3~seconds of the recording. All audio segments were standardized to 3~seconds via zero-padding or truncation, yielding 150,557 preprocessed 3-second clips.

A \textit{no\_bird} class was constructed from low-energy segments of bird recordings and selected ESC-50 environmental sounds. These included natural ambient sounds (e.g., rain, wind, fire), common animal sounds (e.g., dog, cat, frog), and mechanical/environmental sounds (e.g., footsteps, car horn, engine, train). All bird-related and urban classes were explicitly excluded to ensure clear class separation.



\subsection{Training Setup}

We conducted knowledge distillation \cite{hinton2015distilling} experiments using BirdNET-Analyzer as the teacher network to extract soft labels with a confidence threshold of 0.05. Our student architecture employed a Wren-Net featuring either 64 mel-frequency bins, 64 linear filters, and differentiable breakpoint/transition parameters initialized at 8kHz and 200 respectively or a fully learnable filterbank starting from linear bins. Audio segments were processed as 3-second clips at 32kHz sampling rate with 512-point FFT and 320-sample hop length.

The training employed adaptive focal \cite{lin2017focal} distillation loss combining cross-entropy on hard labels with temperature-scaled KL divergence on teacher predictions: $\mathcal{L} = (1-\alpha)\mathcal{L}_{\text{focal}} + \alpha\mathcal{L}_{\text{soft}}$ where $\alpha=0.4$, temperature $T=3.0$, and focal parameter $\gamma=4.0$. Automatic class weighting addressed data imbalance, with adaptive mechanisms adjusting distillation strength based on teacher confidence (adaptation rate 0.1). We used AdamW optimization with learning rate $10^{-3}$, weight decay 0.01, and cosine annealing scheduling over 150 epochs. Filter parameters received enhanced learning rates (15× for breakpoint, 5× for transition width) with dedicated schedulers and exploration techniques including gradient noise injection, oscillatory perturbations every 5 epochs, and momentum resets to escape local optima.

Training incorporated alternating optimization phases: 5 epochs of joint training, followed by cycles of 5 epochs optimizing main network parameters and 1 epoch for filter-specific optimization, concluding with 20 epochs of joint refinement. Early stopping used 35-epoch patience with validation feedback guiding filter parameter search within 10\% ranges. Batch size of 64 with standard data augmentation completed the protocol.
\vspace{-2mm}
\section{Results}

\begin{table}[!ht]
    \centering
    \small
    \renewcommand{\arraystretch}{1}
    \begin{adjustbox}{width=\linewidth,center}
    \begin{tabular}{lcccc}
        \textbf{Filterbank Type} & \textbf{Learning Mode} & \textbf{Best Val Acc (\%)} & \textbf{Test Acc (\%)} \\ \hline
              Mel                      & Fixed                   & 82.49                      & 79.61 \\
        Linear Triangular        & Fixed                   & 82.10                      & 81.45 \\
  
        Combined Log Linear      & Semi-learnable          & \textbf{85.74}                      & \textbf{87.22} \\
        Fully Learnable         & Fully learnable         & 84.26                      & 83.83 \\ \hline
    \end{tabular}
    \end{adjustbox}
    \caption{Comparison of different filterbank approaches (all using 64 bins) on 9 bird species: \textit{Poecile montanus}, \textit{Certhia familiaris}, \textit{Apus apus}, \textit{Bubo bubo}, \textit{Periparus ater}, \textit{Emberiza cia}, \textit{Lophophanes cristatus}, \textit{Certhia brachydactyla}, and the \textit{no\_bird} class.}
    \label{tab:filterbank_comparison}
\end{table}

We evaluated WrenNet with semi-learnable spectral features on our 70-species alpine dataset (Table~\ref{tab:results}). Single-species classification yielded accuracies of 92--95\%, with learned breakpoints at 1955~Hz for \textit{Corvus corax} and 5269~Hz for \textit{Upupa epops}. These breakpoints define the transition frequency between logarithmic and linear frequency mappings, automatically discovered without manual tuning.
For multi-species classification, we observed performance variations across different species groupings. The eight acoustically distinctive species (ravens, owls, herons, woodpeckers) achieved 90.8\% accuracy with our approach. 
The 13-species subset of closely related thrushes, tits, and kinglets yielded 77.5\% accuracy with a learned breakpoint of 1224~Hz. Within this group, the taxonomically similar \textit{Regulus} pair achieved 85.6\% accuracy with a breakpoint of 1221~Hz.
Frequency-based analysis showed consistent patterns: high-frequency specialists achieved 91.5\% accuracy with a 164~Hz breakpoint, while low-frequency species achieved 91.6\% with a 237~Hz breakpoint. Both configurations approximated linear scaling.
Full 70-species classification resulted in 66.5\% accuracy (57k parameters) and 70.1\% accuracy (136k parameters). The 3.6 percentage point improvement from doubling model capacity was accompanied by a shift in learned breakpoints from 851~Hz to 1390~Hz. As classification complexity increased across all experiments, the learned mappings consistently approached linear scaling characteristics.
Table \ref{tab:filterbank_comparison} instead showcases a comparison between our semi-learnable frequency mapping against linear, mel, and fully learnable filterbanks. The fully learnable filterbank is initialized as a linear filterbank and learnt through backpropagation.
\subsection{On Device benchmark}

To evaluate the practical deployment feasibility of our approach, we conducted on-device benchmarks measuring energy consumption, inference time, and power usage across different hardware platforms. All benchmarks were performed on 3-second audio clips processed in 200ms chunks. For the AudioMoth development board, we developed custom benchmark firmware and converted our TensorFlow Lite model to CMSIS-NN \cite{lai2018cmsisnnefficientneuralnetwork} format, deploying it on the device's external SRAM. The Raspberry Pi 3 B+ experiments utilized the TensorFlow Lite runtime engine. Table~\ref{tab:benchmark} presents the results for three configurations: our optimized CMSIS-NN implementation on AudioMoth, our TensorFlow Lite model on Raspberry Pi 3 B+, and BirdNET (8-bit quantized) as a baseline comparison.
The AudioMoth, being a low-power microcontroller-based platform designed for wildlife monitoring, demonstrates the most energy-efficient performance at 0.077 J per inference, though with longer inference times (1.69 s) due to its limited computational resources, namely 80MHz processor and 8-bit bus for external SRAM. In contrast, the Raspberry Pi 3 B+, a more powerful single-board computer, achieves significantly faster inference (0.061 s) while consuming more energy (0.172 J per inference). This trade-off reflects the different design philosophies: ultra-low power consumption versus computational performance. Compared to BirdNET (2.79 J, 0.978 s) on the same Raspberry Pi hardware, our approach requires 16× less energy per inference  and is 16× faster, highlighting the efficiency gains of our lightweight approach.
Overall, these results demonstrate energy consumption suitable for real world wildlife monitoring applications.

\begin{table}[!ht]
    \centering
    \scriptsize
    \renewcommand{\arraystretch}{1}
    \begin{adjustbox}{width=\linewidth,center}
    \begin{tabular}{lccc}
        \textbf{Device} & \textbf{Energy/Inf. [J]} & \textbf{Time/Inf. [s]} & \textbf{Power [W]} \\ \hline
        Audiomoth         & 0.077 & 1.69  & 0.046 \\
        RPi 3 B+ (ours)   & 0.172 & 0.061 & 2.80  \\
        RPi 3 B+ (Birdnet)& 2.79  & 0.978 & 2.84  \\ \hline
    \end{tabular}
    \end{adjustbox}
    \caption{On-device benchmark results, including the energy needed for one 3s inference (J), the inference time (s) and the Power Usage (W).}
    \label{tab:benchmark}
\end{table}
\vspace{-6mm}
\section{Conclusion and Future Works}

This work presented WrenNet, a novel neural network architecture that enables practical multi-species bird audio classification on low-power microcontrollers. Combined with our semi-learnable spectral feature extractor—an efficient and adaptive alternative to fixed filterbanks—the system achieves 70.1\% accuracy on a targeted 70-species dataset and over 90\% for acoustically distinctive species.
Crucially, we demonstrated a dual-path impact: The WrenNet architecture is the first of its kind to run multi-species classification on a low-power device (AudioMoth), while also serving as a highly efficient alternative for more capable hardware. By reducing the inference energy cost by 16x compared to BirdNET on a Raspberry Pi, our model can significantly extend the battery life of Pi-based deployments, enabling longer-term studies.
Future work will explore the deployment of a distributed system of these nodes to investigate collaborative inference strategies, further enhancing the capabilities of network-wide monitoring.
All scripts for dataset creation, training, distillation and model export are available in a public repository\footnote{https://github.com/wren-framework/wrennet}.

\newpage
\bibliographystyle{IEEEbib}
\bibliography{refs}

\end{document}